\documentclass[english,reprint,amsmaths,amssymb,prl]{revtex4-1}
\usepackage[T1]{fontenc}
\usepackage[latin9]{inputenc}
\usepackage{color}
\usepackage{babel}
\usepackage{array}
\usepackage{refstyle}
\usepackage{textcomp}
\usepackage{multirow}
\usepackage{amstext}
\usepackage{graphicx}
\usepackage[unicode=true,pdfusetitle,
 bookmarks=false,
 breaklinks=false,pdfborder={0 0 0},pdfborderstyle={},backref=false,colorlinks=true]
 {hyperref}
\hypersetup{
 pdfborderstyle={}}

\makeatletter

\AtBeginDocument{\providecommand\figref[1]{\ref{fig:#1}}}
\AtBeginDocument{\providecommand\tabref[1]{\ref{tab:#1}}}
\newcommand{\lyxmathsym}[1]{\ifmmode\begingroup\def\b@ld{bold}
  \text{\ifx\math@version\b@ld\bfseries\fi#1}\endgroup\else#1\fi}

\providecommand{\tabularnewline}{\\}
\newcommand{\lyxdot}{.}

\RS@ifundefined{subsecref}
  {\newref{subsec}{name = \RSsectxt}}
  {}
\RS@ifundefined{thmref}
  {\def\RSthmtxt{theorem~}\newref{thm}{name = \RSthmtxt}}
  {}
\RS@ifundefined{lemref}
  {\def\RSlemtxt{lemma~}\newref{lem}{name = \RSlemtxt}}
  {}

\usepackage{babel}
\renewcommand{\tabref}{\Tabref}
\renewcommand{\figref}{\Figref}
\usepackage[shortcuts]{extdash}

\makeatother

\begin{document}

\title{Magnetism and Interlayer Bonding in Pores of Bernal-Stacked Hexagonal
Boron Nitride}

\author{Mehmet Dogan$^{1,2}$ and Marvin L. Cohen$^{1,2}$}

\affiliation{$^{1}$Department of Physics, University of California, Berkeley,
California 94720, USA $\linebreak$ $^{2}$Materials Sciences Division,
Lawrence Berkeley National Laboratory, Berkeley, California 94720,
USA}
\begin{abstract}
When single-layer \emph{h}-BN is subjected to a high-energy electron
beam, triangular pores with nitrogen edges are formed. Because of
the broken $sp^{2}$ bonds, these pores are known to possess magnetic
states. We report on the magnetism and electronic structure of triangular
pores as a function of their size. Moreover, in the Bernal-stacked
\emph{h}-BN (AB-\emph{h}-BN), multilayer pores with parallel edges
can be created, which is not possible in the commonly fabricated multilayer
$\text{AA}^{\prime}$-\emph{h}-BN. Given that these pores can be manufactured
in a well-controlled fashion using an electron beam, it is important
to understand the interactions of pores in neighboring layers. We
find that in certain configurations, the edges of the neighboring
pores remain open and retain their magnetism, and in others, they
form interlayer bonds. We present a comprehensive report on these
configurations for small nanopores. We find that at low temperatures,
these pores have near degenerate magnetic configurations, and may
be utilized in magnetoresistance and spintronics applications. In
the process of forming larger multilayer nanopores, interlayer bonds
can form, reducing the magnetization. Yet, unbonded parallel multilayer
edges remain available at all sizes. Understanding these pores is
also helpful in a multitude of applications such as DNA sequencing
and quantum emission.
\end{abstract}
\maketitle

\section{Introduction\label{sec:Introduction}}

Low-dimensional materials enable a wide range of uses not possible
with bulk (three-dimensional) materials. Creating holes in two-dimensional
(2D) materials, in particular, allows for applications like molecular
sieving, metamaterials, and quantum emission \citep{koenig2012selective,choi2016engineering,zhao2019etching,caldwell2019photonics,liu2021synthesis,song2021deepultraviolet}.
Researchers can tailor a system for a specific application by controlling
the size, shape, and/or distribution of holes in a 2D material. DNA
sequencing using nanopores (holes as small as a few nm) is one example
of such an application, which would allow for quick and precise sequencing
of single unbroken DNA strands \citep{liu2013boronnitride,zhou2013dnatranslocation,muthukumar2015singlemolecule,heerema2016graphene,deamer2016threedecades}.
Nanopores also spontaneously form in 2D materials, so, understanding
their properties is important in itself. In this study, we investigate
the properties of triangular pores in single-layer hexagonal boron
nitride (\emph{h}-BN) and their interactions with pores in neighboring
layers, which is largely determined by the stacking sequence and pore
alignment. We focus primarily on Bernal-stacked \emph{h}-BN which
can house parallel-edged triangular pores in neighboring layers because
of the lack of rotation between the layers \citep{dogan2020magnetic}.

\begin{figure}
\centering{}\includegraphics[width=0.7\columnwidth]{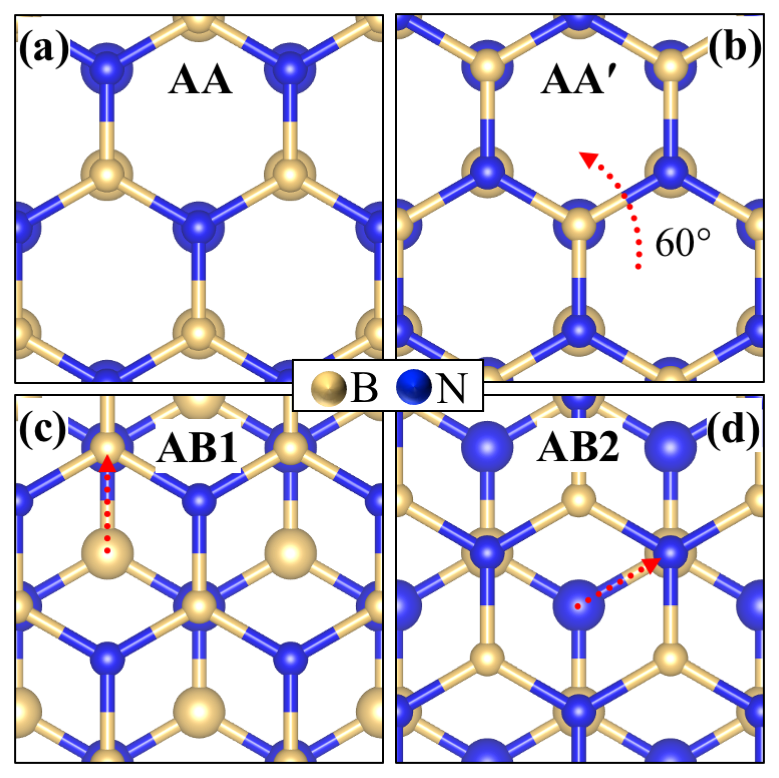}\caption{\label{fig:Stackings}Four high-symmetry stacking sequences of \emph{h}-BN
considered in this study. (a) The AA stacking sequence which has not
been observed experimentally. (b) The commonly observed $\text{AA}^{\prime}$
stacking sequence. (c) The first way of constructing the AB stacking
sequence (AB1) of \emph{h}-BN. (d) The second way of constructing
the AB stacking sequence (AB2) of \emph{h}-BN. The two stacking sequences
in (c,d) are physically equivalent but geometrically distinct. They
are distinguished so that the top and the bottom layers can be treated
separately in this work. The atoms in the bottom layer are enlarged
to ease inspection.}
\end{figure}

Layers of\emph{ h}-BN can be stacked in different ways, resulting
in different material properties \citep{qi2007planarstacking,marom2010stacking,constantinescu2013stacking,gilbert2019alternative}.
The AA stacking {[}\figref{Stackings}(a){]} is a trivial stacking
sequence with no in-plane shift or rotation between successive layers,
and it has an interlayer distance of $3.64\ \mathring{\text{A}}$.
We include this unobserved high-energy stacking sequence in our study
as a reference. The most commonly synthesized stacking sequence of
\emph{h}-BN is the $\text{AA}^{\prime}$ stacking {[}\figref{Stackings}(b){]},
in which each layer is a $60^{\circ}$ rotated copy of the preceding
layer, resulting in columns of alternating B and N atoms in the bulk
\citep{alem2009atomically}. A less common stacking sequence of \emph{h}-BN
is the AB (Bernal) stacking, which includes no relative rotation but
only a relative shift between the layers {[}\figref{Stackings}(c,d){]}.
AB-\emph{h}-BN is energetically very close to $\text{AA}^{\prime}$-\emph{h}-BN
\citep{gilbert2019alternative}, but until recently, it was only observed
in rare cases \citep{warner2010atomicresolution,khan2016carbonand,ji2017chemical}.
The procedure to reliably manufacture this stacking was recently reported
\citep{gilbert2019alternative}, followed by several new studies on
this material \citep{dogan2020electron,yasuda2021stackingengineered,stern2021interfacial,woods2021chargepolarized}.
For a comprehensive comparison of the stacking sequences in \emph{h}-BN,
we refer the reader to Ref. \citep{gilbert2019alternative}.

\begin{figure*}
\centering{}\includegraphics[width=1\textwidth]{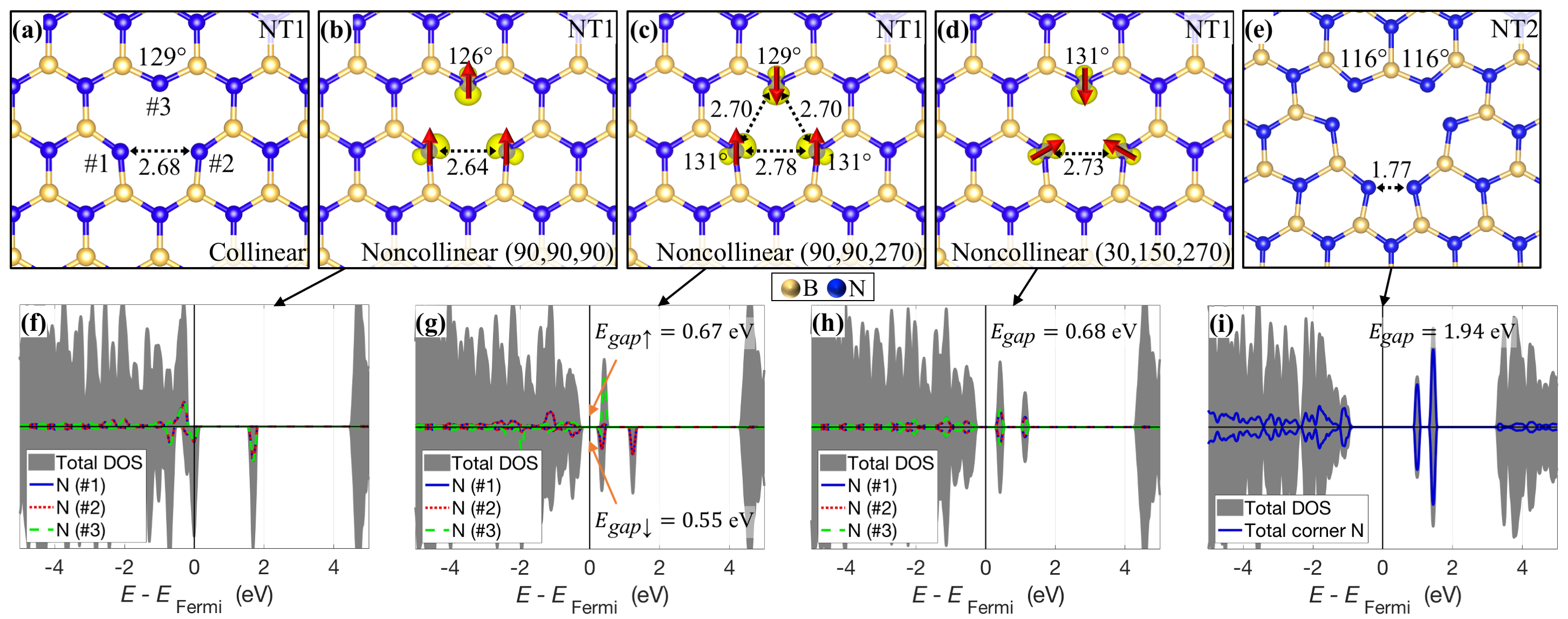}\caption{\label{fig:Monolayer}Atomic structures of the collinear (a) and noncollinear
(b\textendash d) spin configurations in the nitrogen-terminated pore
with 1 nitrogen per edge (NT1); atomic structure of the nitrogen-terminated
pore with 2 nitrogens per edge (NT2); spin-resolved projected densities-of-states
(PDOS) plots for the noncollinear NT1 pores (f\textendash h); PDOS
plot for the NT2 pore (i). The isosurface plots for absolute magnetization
with the isovalue $\left|n_{\uparrow}-n_{\downarrow}\right|=0.02\ \left|e\right|/a_{0}^{3}$
are included in the atomic structure pictures. The direction of magnetization
on each magnetized atom is denoted by a red arrow, and the structures
are labeled by these directions at the bottom right corner of panels
(b\textendash d). The magnetization directions are measured in degrees
from the zigzag direction. The atoms on whose orbitals the densities-of-states
are projected are labeled in (a). The Fermi energy is set to the mid-gap
in cases where there is a band gap.}
\end{figure*}

An infinite sheet of single-layer \emph{h}-BN is a wide-gap insulator,
yet, computational investigations have revealed that its edges and
pores have a diverse set of electronic and magnetic properties \citep{barone2008magnetic,lai2009magnetic,mukherjee2011edgestabilities,yamijala2013electronic,deng2017theedge,dogan2020electron,dogan2020magnetic}.
The unpaired electrons that occupy dangling $sp^{2}$ hybrid orbitals
at edges and pores result in magnetism which is absent in the pristine
sheet. The most commonly formed vacancies in \emph{h}-BN are boron
monovacancies (i.e. a single boron atom removed from the sheet), shown
in \figref{Monolayer}. These vacancies can be created by a high-energy
electron beam in a transmission electron microscopy (TEM) chamber.
As the beam is applied for longer periods of time, the vacancies grow
into larger pores in predictable ways \citep{alem2011vacancy,pham2016formation,gilbert2017fabrication,gilbert2019alternative,dogan2020electron}.
The pores have three nitrogen-terminated zigzag edges that form an
equilateral triangle. In the large pore limit, these edges approximate
the edges of an infinite sheet (N-edge), which is the most common
edge type in \emph{h}-BN \citep{zobelli2007electron,alem2009atomically,meyer2009selective,kotakoski2010electron,kim2011controlled,ryu2015atomicscale,rajan2019addressing,mouhoub2020quantitative},
and has magnetic properties that we elucidated in a previous computational
study \citep{dogan2020magnetic}. In the current study, we first investigate
the spin configurations of nanopores of various sizes in single-layer
\emph{h}-BN and relate them to the infinite edges. We then discuss
the interactions of nested/aligned pores in multilayer \emph{h}-BN.
In the common $\text{AA}^{\prime}$-\emph{h}-BN, consecutive layers
are rotated by $60^{\circ}$ with respect to each other, and triangular
pores do not neatly align. In contrast, multilayer nested pores are
observed in AB-\emph{h}-BN with perfectly parallel edges \citep{gilbert2019alternative,dogan2020electron}.
We report the geometries and electronic structures of all basic bilayer
triangular pore combinations in which the larger of the two pores
is 4 u.c. in size, and generalize the bonding patterns (or lack thereof)
when available.

\begin{table*}
\def\arraystretch{2.0}
\begin{centering}
\begin{tabular}{c|c|c|c|c|c}
\multirow{2}{*}{} & \multicolumn{4}{c|}{NT1 (3)} & NT2 (0)\tabularnewline
\cline{2-6} 
 & Collinear & NC$\left(90,90,90\right)$ & NC$\left(90,90,270\right)$ & NC$\left(30,150,270\right)$ & Collinear\tabularnewline
\hline 
$\Delta E$ (eV) & $\equiv0$ & \multicolumn{1}{c|}{$-0.138$} & $-0.185$ & $-0.187$ & $\equiv0$\tabularnewline
\hline 
ES type & metal & \multicolumn{1}{c|}{half-metal} & magnetic SC & nonmagnetic SC & nonmagnetic SC\tabularnewline
\hline 
$E_{gap\uparrow}$ (eV) & 0 & 4.64 & $0.67$ & $0.68$ & 1.94\tabularnewline
\hline 
$E_{gap\downarrow}$ (eV) & 0 & 0 & $0.55$ & $0.68$ & 1.94\tabularnewline
\end{tabular}
\par\end{centering}
\caption{\label{tab:MLpores}Total energies, electronic state types and band
gaps of the NT1 and NT2 nanopores. Refer to the text for the explanation
of the labels in the second row. The total energies are with reference
to the collinear configuration and presented per edge N atom, for
instance, $\Delta E$ of and NT1 configuration is computed by subtracting
the energy of the collinear configuration from it and dividing by
3 (the number of edge N atoms).}
\end{table*}

We conclude this section by noting that in AB-\emph{h}-BN, the two
consecutive layers are inequivalent. With the same bottom layer, the
top layer may be shifted in two distinct ways to construct the AB
stacking, shown in \figref{Stackings}(c,d) by the red arrows, which
we call AB1 and AB2. In order to keep track of all nanopore combinations,
we distinguish between AB1 and AB2 stackings for the rest of the study.
This property of AB-\emph{h}-BN clearly distinguishes it from $\text{AA}^{\prime}$-\emph{h}-BN
and was recently utilized to generate ferroelectricity in bilayer
\emph{h}-BN \citep{yasuda2021stackingengineered,stern2021interfacial,woods2021chargepolarized}.
For a more detailed discussion of this point, we refer the reader
to Ref. \citep{dogan2020magnetic}.

\section{Methods\label{sec:Methods}}

We conduct density functional theory (DFT) calculations within the
Perdew\textendash Burke\textendash Ernzerhof generalized gradient
approximation (PBE GGA) \citep{perdew1981selfinteraction}, using
the QUANTUM ESPRESSO software package with norm-conserving pseudopotentials
\citep{giannozzi2009quantum,hamann2013optimized}. The plane-wave
energy cutoff for the pseudo Kohn-Sham wavefunctions used is 80 Ry.
For a $1\times1$ unit cell of \emph{h}-BN, we use a $12\times12\times1$
Monkhorst\textendash Pack k-point mesh to sample the Brillouin zone
\citep{cohen1975selfconsistent}, and choose k-point meshes accordingly
for larger unit cells. A $\sim$14 Å of vacuum is placed between the
copies of the 2D system along the out-of-plane direction to isolate
the sheets. In order to include the interlayer van der Waals interactions,
we include a Grimme-type dispersion correction \citep{grimme2006semiempirical}.
All atomic coordinates are relaxed until the forces on all the atoms
are less than $10^{-3}$ Ry/$a_{0}$ in all three Cartesian directions,
where $a_{0}$ is the Bohr radius.

\section{Results\label{sec:Results}}

\subsection{Nanopores in monolayer \emph{h-}BN\label{subsec:Monolayer}}

We start with a survey of the possible spin configurations in the
boron monovacancy in the single layer \emph{h}-BN\emph{.} Throughout
this article, we label pores as ``NT$n$'', where ``NT'' stands
for ``nitrogen-terminated'', and $n$ is the number of nitrogen
atoms on each edge of the triangle. Thus, a boron monovacancy is denoted
as NT1. In \figref{Monolayer}(a-d), we present the atomic structure
of NT1 in its various magnetic configurations. \figref{Monolayer}(a)
is obtained by running a collinear calculation, resulting in a structure
with 3-fold symmetry. The structures in \figref{Monolayer}(b-d) are
obtained by relaxing the atoms further with noncollinear spin channels
and different starting configurations. All three of these configurations
have lower energies compared to the collinear structure, as listed
in \tabref{MLpores}, and are local minima in the configuration space.
We find that the 3-fold symmetric NC(30,150,270) structure has the
lowest energy, but only by $\sim2$ meV, compared to NC(90,90,270).
Interestingly, the latter configuration is a triplet of atomic structures,
where there are two shorter N\textendash N distances (2.70 $\mathring{\text{A}}$)
and a longer N\textendash N distance (2.78 $\mathring{\text{A}}$).
At regular temperatures (20 K $<T<$ 500 K), we expect both the NC(30,150,270)
and NC(90,90,270) pores to be present in single layer \emph{h}-BN
with pores with only trace amounts of NC(90,90,90). To the best of
our knowledge, previous studies have only identified NC(90,90,90)
\citep{si2007magnetic,du2009dotsversus,zhao2018magnetism,dogan2020electron}.
Although no probing of the local magnetism of edges or pores in \emph{h}-BN
has been reported, ferromagnetism has been shown to occur in \emph{h}-BN
at room temperature \citep{si2014intrinsic}, which can be attributed
to the existence of edges and pores in regular \emph{h}-BN samples.

The next smallest triangular pore is NT2, which is obtained by removing
three borons and one nitrogen from the system (\figref{Monolayer}(e)).
In this pore, the corner nitrogens approach each other and form dimers,
resulting in a collinear electronic structure ($\left|n_{\uparrow}-n_{\downarrow}\right|=0$
everywhere), as reported previously \citep{du2009dotsversus,dogan2020electron}.

Looking at the electronic structures of the NT1 pores (\figref{Monolayer}(f-h)),
we observe that the NC(90,90,90) pore is a half-metal, i.e. the majority
spin channel is insulating and the minority spin channel is a metal,
the NC(90,90,270) pore is a magnetic semiconductor and the NC(30,150,270)
pore is a nonmagnetic semiconductor, although the electronic structures
of the latter two configurations are very similar. The antiferromagnetic
state is a semiconductor with a gap of 0.51 eV. Because the spin configuration
and the electronic structure are closely linked, these pores may be
useful in various spintronics applications\textcolor{black}{{} \citep{prinz1998magnetoelectronics,wolf2001spintronics,vzutic2004spintronics,awschalom2007challenges,felser2007spintronics,yazyev2010emergence,awschalom2013quantum,ahn20202dmaterials}.}
As seen in \figref{Monolayer}(i), the electronic structure of the
NT2 pore is a nonmagnetic semiconductor with a larger gap (\tabref{MLpores}).
We present the atomic and electronic structures of larger pores in
single layer \emph{h}-BN (NT3\textendash NT8) in Figures S1\textendash S6.
The key information for these larger pores is summarized in \tabref{MLsize}.
In each case, we fully relax the pore in a collinear calculation,
and then run a self-consistent field calculation with two initial
spin configurations: where all the edge spins are set as parallel
(P), and where consecutive edge spins are set as antiparallel (AP).
We note that we have confirmed in a few test cases that in these larger
pores, further relaxations with noncollinearity cause negligible changes
in atomic positions. In Figures S1\textendash S6, the magnetization
in the real space is only plotted for the P configurations (the AP
configurations have equivalent plots with alternating sign). In \tabref{MLsize},
we also include the case of the N-edge taken from Ref. \citep{dogan2020magnetic},
which may be interpreted as the limiting case NT$\infty$.

From the results in Figures S1\textendash S6 and \tabref{MLsize},
we make a few observations: (i) The corner N atoms remain dimerized
in these larger pores and thus have no magnetic moment. This causes
the sequence of the absolute value of the magnetic moments to be approximately
3$\mu_{\mathrm{B}}$, 0, 3$\mu_{\mathrm{B}}$, 6$\mu_{\mathrm{B}}$,$...$
for the NT1, NT2, NT3, NT4,$...$ pores. (ii) For the NT$n$ pores
with $n\geq2$, the energy differences between P and AP spin configurations
is smaller compared to NT1 and the N-edge. The comparison with NT1
can be explained by the isolation provided by the dimerized corner
N atoms. The comparison with the N-edge may be due to the fact that
these pores are still small, and many of the edge N atoms lack nearest
and next-nearest neighbors to interact with, as opposed to the infinite
edges (we know that both nearest and next-nearest neighbor interactions
are important for the spin dynamics in this system, cf. \citep{dogan2020magnetic}).
This near degeneracy of different spin configurations would be advantageous
in applications that rely on magnetic switching. (iii) The P configurations
are consistently half-metals, and the AP configurations are either
magnetic semiconductors (MSC) or nonmagnetic semiconductors (NMSC),
depending on the number of available magnetic sites (written in parentheses
in the first row of \tabref{MLpores} and the first column of \tabref{MLsize}).
The band gap of the NMSC configurations gradually increases with the
pore size, approaching the value of the N-edge (0.51 eV). (iv) The
positions of the in-gap states for the AP configurations depend on
the pore size. Because the pore size can be controlled by the duration
of the applied electron beam \citep{pham2016formation,gilbert2017fabrication},
it should be possible to engineer the desired in-gap states at a desired
location on the \emph{h}-BN layer.

\begin{table}
\def\arraystretch{2.0}
\begin{centering}
\begin{tabular}{c|c|c|c|c}
 &  & $\Delta E$ (eV) & ES type & $E_{gap}$ (eV)\tabularnewline
\hline 
\multirow{2}{*}{NT3 (3)} & P & $-0.230$ & half-metal & $0.41,0$\tabularnewline
\cline{2-5} 
 & AP & \multicolumn{1}{c|}{$-0.229$} & MSC & $0.15,0.09$\tabularnewline
\hline 
\multirow{2}{*}{NT4 (6)} & P & \multicolumn{1}{c|}{$-0.235$} & half-metal & $0.46,0$\tabularnewline
\cline{2-5} 
 & AP & $-0.242$ & NMSC & $0.31,0.31$\tabularnewline
\hline 
\multirow{2}{*}{NT5 (9)} & P & $-0.220$ & half-metal & $0.58,0$\tabularnewline
\cline{2-5} 
 & AP & $-0.225$ & MSC & $0.18,0.29$\tabularnewline
\hline 
\multirow{2}{*}{NT6 (12)} & P & $-0.225$ & half-metal & $0.70,0$\tabularnewline
\cline{2-5} 
 & AP & $-0.236$ & NMSC & $0.33,0.33$\tabularnewline
\hline 
\multirow{2}{*}{NT7 (15)} & P & $-0.226$ & half-metal & $0.72,0$\tabularnewline
\cline{2-5} 
 & AP & $-0.227$ & MSC & $0.25,0.41$\tabularnewline
\hline 
\multirow{2}{*}{NT8 (18)} & P & $-0.226$ & half-metal & $0.95,0$\tabularnewline
\cline{2-5} 
 & AP & $-0.229$ & NMSC & $0.36,0.36$\tabularnewline
\hline 
\multirow{2}{*}{N-edge \citep{dogan2020magnetic}} & P & $-0.218$ & half-metal & $4.56,0$\tabularnewline
\cline{2-5} 
 & AP & $-0.184$ & NMSC & $0.51,0.51$\tabularnewline
\end{tabular}
\par\end{centering}
\caption{\label{tab:MLsize}Total energies, electronic state types and band
gaps of the NT3\textendash NT8 nanopores. Refer to the text for the
explanation of P and AP labels. The total energies are with reference
to the collinear configuration and presented per edge N atom, for
instance, $\Delta E$ of an NT3 configuration is computed by subtracting
the energy of the collinear configuration from it and dividing by
9 (the number of edge N atoms). The numbers in parentheses in the
first column denote the number of magnetic N atoms for each pore.}
\end{table}

\subsection{Nanopores in bilayer \emph{h}-BN}

When nanopores are created on multilayer \emph{h}-BN, they tend to
begin on the top layer and then propagate into the layers underneath
\citep{gilbert2017fabrication,dogan2020electron}. As a first step
in understanding nanopores in multilayer systems, we present results
on nanopores on the top layer of bilayer \emph{h}-BN, where the bottom
layer remains as a full sheet. In Figures S7\textendash S22, we present
the atomic and electronic structures of pores NT1\textendash NT4 in
a single sheet stacked on a full sheet using AA, $\text{AA}^{\prime}$,
AB1 and AB2 stacking sequences. For NT1, NT3 and NT4 pores, parallel
(P) and antiparallel (AP) spin configurations are run without further
relaxation of the atomic positions (NT2 remains fully spin-unplorazied),
and the magnetization in the real space is only plotted for the P
configurations (the AP configurations have equivalent plots with alternating
sign). Our findings are also summarized in \tabref{MLporeBL}. In
general, the behavior of the pores in the bilayer remains very similar
to their behavior in single \emph{h}-BN sheets, although the small
shifts in the energy levels. However, the magnetized regions around
the edge N atoms do not lie entirely parallel to the sheet for the
larger pores (NT3 and up), instead, these spin-polarized dangling
orbitals point away from the bottom sheet by an angle $\theta$, which
is listed for each case in the table. 

\begin{table*}
\def\arraystretch{2.0}
\begin{centering}
\begin{tabular}{c|c|c|c|c|c|c|c|c|c|c|c|c|c}
\multicolumn{1}{c}{} &  & \multicolumn{3}{c|}{$\text{AA}$} & \multicolumn{3}{c|}{$\text{AA}^{\prime}$} & \multicolumn{3}{c|}{$\text{AB1}$} & \multicolumn{3}{c}{$\text{AB2}$}\tabularnewline
\cline{3-14} 
\multicolumn{1}{c}{} &  & $\Delta E$ (eV) & $E_{gap}$ (eV) & $\theta$ & $\Delta E$  & $E_{gap}$  & $\theta$ & $\Delta E$  & $E_{gap}$  & $\theta$ & $\Delta E$  & $E_{gap}$  & $\theta$\tabularnewline
\hline 
\multirow{2}{*}{NT1 (3)} & P & $-0.157$ & $4.08,0$ & $0\lyxmathsym{\textdegree}$ & $-0.164$ & $4.36,0$ & $0\lyxmathsym{\textdegree}$ & $-0.166$ & $4.48,0$ & $0\lyxmathsym{\textdegree}$ & $-0.161$ & $4.39,0$ & $0\lyxmathsym{\textdegree}$\tabularnewline
\cline{2-14} 
 & AP & \multicolumn{1}{c|}{$-0.181$} & $0.29,0.18$ & $0\lyxmathsym{\textdegree}$ & $-0.203$ & $0.45,0.34$ & $0\lyxmathsym{\textdegree}$ & $-0.204$ & $0.40,0.29$ & $0\lyxmathsym{\textdegree}$ & $-0.212$ & $0.49,0.38$ & $0\lyxmathsym{\textdegree}$\tabularnewline
\hline 
NT2 (0) & Collinear & $\equiv0$ & 1.75 & n/a & $\equiv0$ & 1.82 & n/a & $\equiv0$ & 1.71 & n/a & $\equiv0$ & 1.86 & n/a\tabularnewline
\hline 
\multirow{2}{*}{NT3 (3)} & P & $-0.224$ & $0.52,0.23$ & $38\lyxmathsym{\textdegree}$ & $-0.228$ & $0.51,0$ & $34\lyxmathsym{\textdegree}$ & $-0.231$ & $0.46,0.20$ & $37\lyxmathsym{\textdegree}$ & $-0.230$ & $0.59,0.08$ & $35\lyxmathsym{\textdegree}$\tabularnewline
\cline{2-14} 
 & AP & $-0.215$ & $0.19,0.17$ & $44\lyxmathsym{\textdegree}$ & $-0.219$ & $0.06,0$ & $16\lyxmathsym{\textdegree}$ & $-0.227$ & $0.21,0.20$ & $38\lyxmathsym{\textdegree}$ & $-0.221$ & $0.07,0$ & $36\lyxmathsym{\textdegree}$\tabularnewline
\hline 
\multirow{2}{*}{NT4 (6)} & P & $-0.232$ & $0.66,0$ & $39\lyxmathsym{\textdegree}$ & $-0.234$ & $0.66,0$ & $39\lyxmathsym{\textdegree}$ & $-0.238$ & $0.61,0.09$ & $39\lyxmathsym{\textdegree}$ & $-0.235$ & $0.58,0.08$ & $37\lyxmathsym{\textdegree}$\tabularnewline
\cline{2-14} 
 & AP & $-0.215$ & $0.31,0.31$ & $42\lyxmathsym{\textdegree}$ & $-0.221$ & $0.22,0.22$ & $6\lyxmathsym{\textdegree}$ & $-0.226$ & $0.26,0.26$ & $39\lyxmathsym{\textdegree}$ & $-0.225$ & $0.29,0.29$ & $5\lyxmathsym{\textdegree}$\tabularnewline
\end{tabular}
\par\end{centering}
\caption{\label{tab:MLporeBL}Total energies, electronic state types, band
gaps and tilt angles of the magnetized regions of the NT1\textendash NT4
nanopores on the top sheet of \emph{h}-BN bilayers. Refer to the text
for the explanation of P and AP labels. The total energies are with
reference to the collinear configuration and presented per edge N
atom, for instance, $\Delta E$ of an NT3 configuration is computed
by subtracting the energy of the collinear configuration from it and
dividing by 9 (the number of edge N atoms). The numbers in parentheses
in the first column denote the number of magnetic N atoms for each
pore.}
\end{table*}

Once a nanopore begins to form in the top layer of multilayer \emph{h}-BN,
the second layer becomes exposed the electron beam, giving rise to
progressively larger and deeper (in terms of the number of layers)
pores \citep{gilbert2017fabrication,dogan2020electron}. In the next
part of our work, we investigate the coexistence of small pores in
both layers of bilayer \emph{h}-BN in order to elucidate the early
stages of this process. In Figures S23\textendash S24, we present
the NT1 pores on both layers of bilayer \emph{h}-BN (in this case,
AB2 stacking is identical to AB1 so it is not presented separately).
The NT1 \& NT1 configurations are uncoupled in the sense that the
pores in the neighboring layers do not interact beyond the already
present van der Waals forces. We note here that for the NT1 \& NT1
configurations, we have checked for potential interlayer coupling
of spins by running a comprehensive collection of spin polarization
combinations. We have not found a noteworthy difference in energies
beyond a simple summation of the energies of spin configurations in
each layer.

As the next step in the process for the AB stacking, we remove the
3 N atoms of the bottom layer that are exposed in addition to the
1N atom on the top layer. This results in the configuration in Figure
S25, which is an unusual defect structure where three B atoms in the
bottom layer move inward within the plane and form a trimer. This
magnetic semiconductor has the band gaps $\left(0.59,0.43\right)$
eV in the two spin channels. We note that this structure is not an
artifact of symmetry and seems to be robust with respect to atomic
position perturbations.

\begin{figure*}
\centering{}\includegraphics[width=0.9\textwidth]{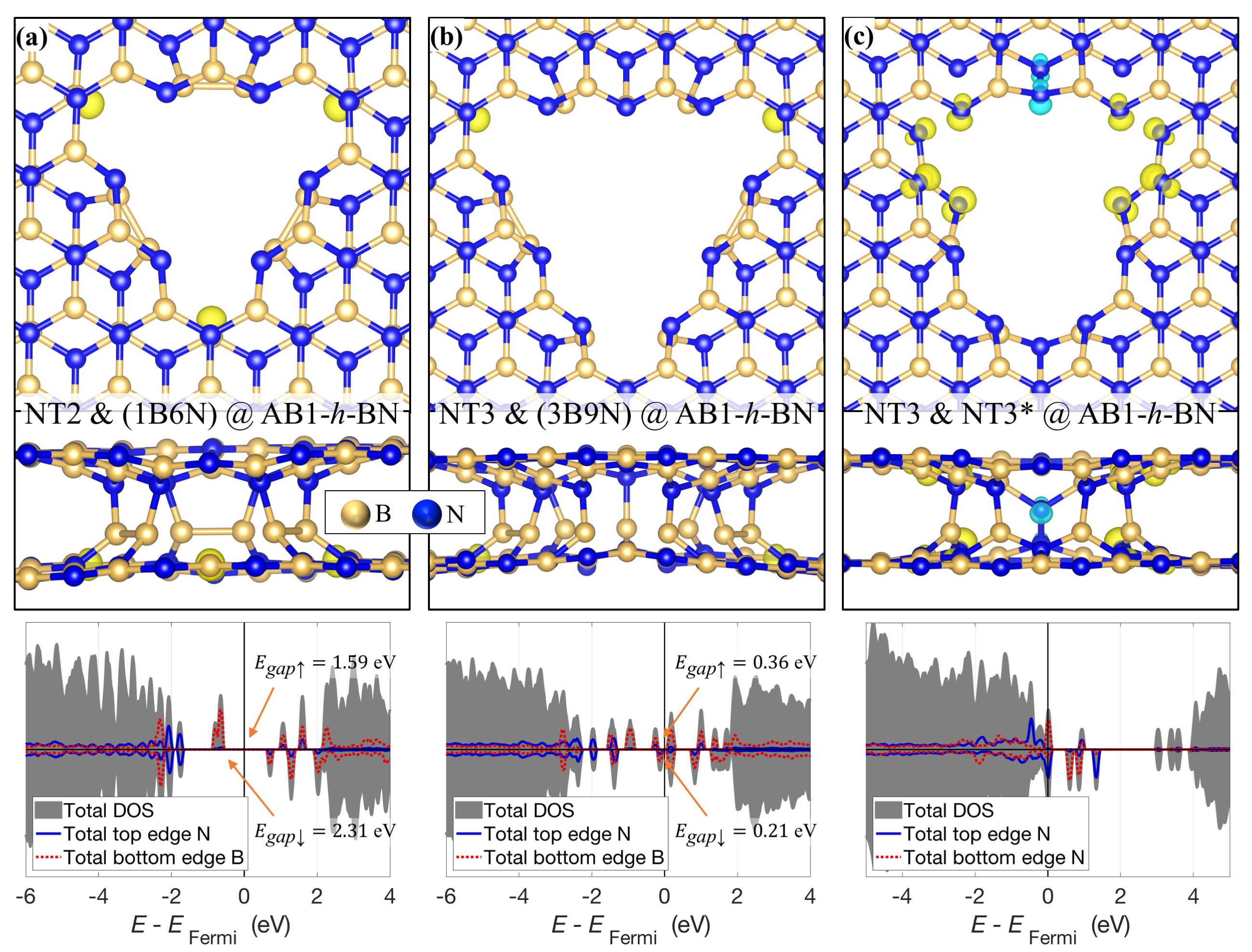}\caption{\label{fig:BLpores_1}Atomic and electronic structures of bilayer
pores in AB1-\emph{h}-BN that result in coupled structures where the
top layer is an NT2 or NT3 pore. (a) The NT2 \& (1B6N) configuration.
(b) The NT3 \& (3B9N) configuration. (c) The NT3 \& NT3{*} configuration.
The isosurface plots for magnetization with the isovalue $n_{\uparrow}-n_{\downarrow}=\pm0.02\ \left|e\right|/a_{0}^{3}$
are included in the atomic structure pictures. Top view and side view
of each nanopore are stacked vertically in each panel. For the spin-resolved
density of states plots in the bottom row, the Fermi energy is set
to the mid-gap in cases where there is a band gap.}
\end{figure*}

For larger pores in bilayer \emph{h}-BN,\emph{ }we focus on the AB
stacking. As discussed above, due to the relative $60\lyxmathsym{\textdegree}$
rotation between consecutive layers in the $\text{AA}^{\prime}$ stacking,
multilayer pores become irregularly shaped, whereas the AB stacking
allows for nested and aligned pores \citep{dogan2020electron}. As
the next step in AB1-\emph{h}-BN, we assume that the NT2 pore has
formed in the top layer, and then we remove the boron in the center
of the exposed area in the bottom layer (cf. Figure S13). This results
in the NT2 \& NT1 structure presented in Figure S26, which is also
a uncoupled structure (half-metal, $E_{gap\uparrow}=1.95$ eV). Removing
the 6 exposed nitrogens from this configuration results in our first
coupled structure, presented in \figref{BLpores_1}(a). This 3-fold
symmetric structure is a magnetic semiconductor which has 6 interlayer
B\textendash N bonds. Forming the NT2 \& NT1 structure in AB2-\emph{h}-BN
is possible in two different ways by removing a boron from either
the center or the corner of the exposed area in the bottom layer (cf.
Figure S14). In both cases, we get uncoupled structures that are presented
in Figures S27\textendash S28 (both magnetic semiconductors with the
band gaps $\left(0.46,0.44\right)$ eV and $\left(0.49,0.26\right)$
eV, respectively). Finally, we present the NT2 \& NT2 structures in
the AA and AB1 stacking sequences in Figures S29\textendash S30, both
of which are uncoupled and nonmagnetic semiconductors with the band
gaps $1.53$ eV and $1.74$ eV, respectively (AB2 is equivalent to
AB1).

We then move to the NT3 pore in the top layer of AB1-\emph{h}-BN.
Removing one of the three exposed borons in the bottom layer (cf.
Figure S17) results in the NT3 \& NT1 uncoupled structure (Figure
S31), which is a half-metal with $E_{gap\uparrow}=0.64$ eV. Similarly,
the NT3 \& NT2 structure in AB1-\emph{h}-BN is also uncoupled (Figure
S32) but metallic. Next, we remove the remaining exposed nitrogens
so that all the atoms in the bottom layer that are exposed when the
top layer has an NT3 pore are removed. This results in the structure
in \figref{BLpores_1}(b) which has reflection symmetry and 9 interlayer
B\textendash N bonds, and is a magnetic semiconductor. An interesting
effect occurs when the NT3 \& NT3 structure in AB1-\emph{h}-BN is
set up and relaxed, in which two edge nitrogens in the bottom layer
detach from their neighbors and form an $\textrm{N}_{2}$ molecule.
When the molecule is taken out of the system, the resulting coupled
structure, which is metallic, is presented in \figref{BLpores_1}(c).
Because the bottom layer has lost two nitrogens, we call this configuration
NT3 \& NT3{*}, and it has 4 interlayer B\textendash N bonds as well
as 1 interlayer N\textendash N bond. This structure demonstrates that
not all nested NT$n_{1}$ \& NT$n_{2}$ pores are possible in their
expected stoichiometry in bilayer and therefore multilayer AB-\emph{h}-BN.

There are three inequivalent NT3 \& NT1 structures in AB2-\emph{h}-BN
depending on which boron in the bottom layer is removed (cf. Figure
S18). These three result in uncoupled half-metallic structures that
are presented in Figures S33\textendash S35 with $E_{gap\uparrow}=0.63,0.64,0.66$
eV, respectively. The two inequivalent NT3 \& NT2 structures in AB2-\emph{h}-BN
are also uncoupled and half-metallic, and presented in Figures S36\textendash S37
($E_{gap\uparrow}=0.63,0.61$ eV, respectively). The NT3 \& NT3 structure
in AB2-\emph{h}-BN is equivalent to that in AB1-\emph{h}-BN (\figref{BLpores_1}(c)).
Finally, we present the NT3 \& NT3 structure in the AA stacking sequence
in Figure S38, which is a magnetic metal with interlayer bonding of
the mid-edge nitrogens.

\begin{figure*}
\centering{}\includegraphics[width=0.9\textwidth]{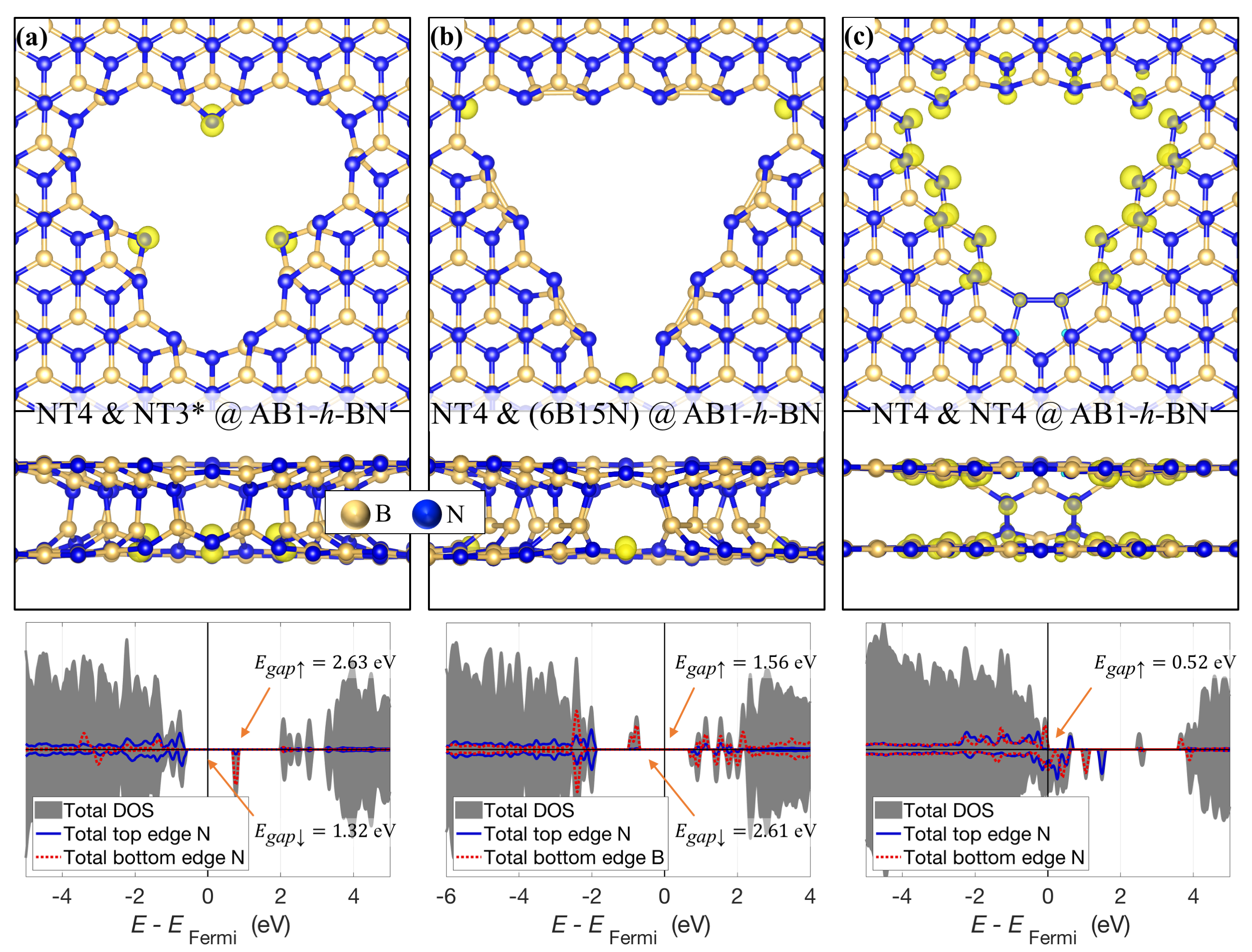}\caption{\label{fig:BLpores_2}Atomic and electronic structures of bilayer
pores in AB1-\emph{h}-BN that result in coupled structures where the
top layer is an NT4 pore. (a) The NT4 \& NT3{*} configuration. (b)
The NT4 \& (6B15N) configuration. (c) The NT4 \& NT4 configuration.
The isosurface plots for magnetization with the isovalue $n_{\uparrow}-n_{\downarrow}=\pm0.02\ \left|e\right|/a_{0}^{3}$
are included in the atomic structure pictures. Top view and side view
of each nanopore are stacked vertically in each panel. For the spin-resolved
density of states plots in the bottom row, the Fermi energy is set
to the mid-gap in cases where there is a band gap.}
\end{figure*}

The largest size we have considered for bilayer \emph{h}-BN in this
study is the NT4 pore. We start with the NT4 \& NT1 configuration
in AB1-\emph{h}-BN of which there are two (Figures S39\textendash S40).
Both of these are uncoupled and half-metallic with $E_{gap\uparrow}=0.66,0.69$
eV, respectively. When the bottom layer has an NT2 pore, interlayer
bonds form, but not all the top edge nitrogens are bonded. This coupled
configuration (NT4 \& NT2) is presented in Figure S41 and is a half-metal
with $E_{gap\uparrow}=0.24$ eV. Enlarging the bottom layer pore one
more step to set up the NT4 \& NT3 structure results in the detachment
of three $\textrm{N}_{2}$ molecules and 12 interlayer B\textendash N
bonds (\figref{BLpores_2}(a)). By losing most of its dangling orbitals,
this system, which we call NT4 \& NT3{*}, becomes a fairly large-gap
magnetic semiconductor. Next, we remove the remaining exposed nitrogens
from this configuration so that all the atoms in the bottom layer
that are exposed when the top layer has an NT4 pore are removed. This
results in the structure in \figref{BLpores_2}(b) which has 12 interlayer
B\textendash N bonds and is a magnetic semiconductor. The NT4 \& NT4
bilayer pore in AB1-\emph{h}-BN also results in some interlayer bonding(\figref{BLpores_2}(c)).
In this configuration, the alignment of the two edges (diagonal in
\figref{BLpores_2}(b)) is different from the alignment of the remaining
edge (horizontal in \figref{BLpores_2}(c). The interlayer bonds form
only at the latter edge. These two edge types correspond to the infinite
edges in Figures 5 and 6 of \citep{dogan2020magnetic}, respectively.
This particular pattern of interlayer bonding should hold for all
NT$n$ \& NT$n$ pores ($n\geq4$) of AB-\emph{h}-BN, where $n-2$
interlayer N\textendash N bonds occur only at one of the edges. 

In AB2-\emph{h}-BN, there are four inequivalent NT4 \& NT1 structures
depending on which boron in the bottom layer is removed (cf. Figure
S22). These structures are presented in Figures S42\textendash S45.
The NT4 \& NT1 (1) structure has a single interlayer N\textendash N
bond and is a magnetic metal. The NT4 \& NT1 (2) structure has two
interlayer N\textendash N bonds and is a magnetic half-metal with
$E_{gap\uparrow}=0.57$ eV. The remaining two NT4 \& NT1 structures
are uncoupled half-metals with $E_{gap\uparrow}=0.48,0.60$ eV, respectively.
The three inequivalent NT4 \& NT2 structures are presented in Figures
S46\textendash S48. These structures are uncoupled half-metals with
$E_{gap\uparrow}=0.57,0.62,0.63$ eV, respectively. Among the two
inequivalent NT4 \& NT3 configurations, the first one (Figure S49)
has 2 interlayer N\textendash N bonds and is a magnetic metal. This
bonding pattern should hold for all NT$n$ \& NT$\left(n-1\right)$
pores ($n\geq4$) of AB-\emph{h}-BN, where $n-3$ interlayer N\textendash N
bonds each occur at two of the edges. The second NT4 \& NT3 configuration
in AB2-\emph{h}-BN (Figure S50) is an uncoupled structure that is
a half-metal with $E_{gap\uparrow}=0.58$ eV. The NT4 \& NT4 structure
in AB2-\emph{h}-BN is equivalent to that in AB1-\emph{h}-BN (\figref{BLpores_2}(c)).
Finally, the NT4 \& NT4 structure in the AA stacking sequence is presented
in Figure S51, which is a magnetic half-metal ($E_{gap\uparrow}=2.78$
eV) with interlayer bonding of the two mid-edge nitrogens on each
edge.

\section{Conclusion\label{sec:Conclusion4}}

We conducted a first-principles study of the nitrogen-terminated triangular
nanopores in \emph{h}-BN with a special focus on Bernal-stacked \emph{h}-BN
(AB-\emph{h}-BN). We found that the ground state of the smallest pore
(B monovacancy) is a 3-fold symmetric configuration with zero total
magnetization. However, the existence of magnetic configurations that
are almost degenerate with the ground state heightens the potential
of magnetic switchability. For larger pores in single-layer \emph{h}-BN,
configurations with parallel and antiparallel neighboring magnetizations
are also close in energy, indicating the likelihood of easier magnetization
of these pores compared to longer edges where energy differences are
larger. For bilayer configurations with pores in both layers, we find
that in most cases, the layers remain uncoupled, \emph{i.e.}, no significant
out-of-plane movement or bonding occurs. However, several structures
that are expected to occur during electron irradiation do form interlayer
bonds, reducing the magnetic character of the pores. We expect these
types of coupled structures to occur during electron irradiation alongside
the simpler decoupled structures. We also find that some nested pores
are not possible in their expected stoichiometry in bilayer and therefore
multilayer AB-\emph{h}-BN as a result of to the detachment of nitrogens
by forming $\textrm{N}_{2}$ molecules. We hope that our results will
motivate experimental studies that closely investigate the magnetic
and electronic properties of these pores in \emph{h}-BN, given their
enormous potential in applications such as DNA sequencing.

\section*{Acknowledgments}

This work was supported by the Director, Office of Science, Office
of Basic Energy Sciences, Materials Sciences and Engineering Division,
of the U.S. Department of Energy under contract No. DE-AC02-05-CH11231,
within the Theory of Materials program (KC2301), which supported first-principles
computations of the atomic structures. Further support for theoretical
work was provided by the NSF Grant No. DMR-1926004 which supported
first-principles computations of the precise electronic structures.
Computational resources used were Cori at National Energy Research
Scientific Computing Center (NERSC), which is supported by the Office
of Science of the US Department of Energy under contract no. DE-AC02-05-CH11231,
Stampede2 at the Texas Advanced Computing Center (TACC) through Extreme
Science and Engineering Discovery Environment (XSEDE), which is supported
by National Science Foundation (NSF) under grant no. ACI-1053575,
Frontera at TACC, which is supported by NSF grant no. OAC-1818253,
and Bridges-2 at the Pittsburgh Supercomputing Center (PSC), which
is supported by NSF award number ACI-1928147.

\bibliographystyle{apsrev4-1}
\bibliography{BN_pores}

\end{document}